 \documentclass[aps,pra,reprint,superscriptaddress,longbibliography]{revtex4-1}

\usepackage{graphicx}
\usepackage{amsmath}

\begin{document}

\title[Resonance behavior of a compliant piezo-driven inkjet channel with an entrained microbubble]{Resonance behavior of a compliant piezo-driven inkjet channel with an entrained microbubble}
\author{Hans Reinten}
\affiliation{Canon Production Printing, P.O. Box 101, 5900 MA Venlo, Netherlands}
\author{Yogesh Jethani}
\author{Arjan Fraters}
\affiliation{Physics of Fluids group, Max-Planck Center Twente for Complex Fluid Dynamics, MESA+ Institute for Nanotechnology, and J. M. Burgers Centre for Fluid Dynamics, University of Twente, P.O. Box 217, 7500 AE Enschede, Netherlands}
\author{Roger Jeurissen}
\affiliation{Department of Applied Physics, Eindhoven University of Technology, P.O. Box 513, 5600 MB Eindhoven, Netherlands}
\author{Detlef Lohse}
\author{Michel Versluis}
\affiliation{Physics of Fluids group, Max-Planck Center Twente for Complex Fluid Dynamics, MESA+ Institute for Nanotechnology, and J. M. Burgers Centre for Fluid Dynamics, University of Twente, P.O. Box 217, 7500 AE Enschede, Netherlands}
\author{Tim Segers}
\affiliation{BIOS Lab-on-a-Chip group, Max-Planck Center Twente for Complex Fluid Dynamics, MESA+ Institute for Nanotechnology, University of Twente, Enschede, Netherlands}
\email{t.j.segers@utwente.nl}

\preprint{Author, JASA}		

\date{\today} 

\begin{abstract}


Microbubbles entrained in a piezo-driven drop-on-demand (DOD) printhead disturb the acoustics of the microfluidic ink channel and thereby the jetting behavior. Here, the resonance behavior of an ink channel as a function of the microbubble size and the number of bubbles is studied through theoretical modeling and experiments. The system is modeled as a set of two coupled harmonic oscillators: one corresponding to the compliant ink channel and one to the microbubble. The predicted and measured eigenfrequencies are in excellent agreement. It was found that the resonance frequency is independent of the bubble size as long as the compliance of the bubble dominates over that of the piezo actuator. An accurate description of the eigenfrequency of the coupled system requires the inclusion of the increased inertance of the entrained microbubble due to confinement. We show that the inertance of a confined bubble can be accurately obtained by using a simple potential flow approach. The model is further validated by the excellent agreement between the modeled and measured microbubble resonance curves. The present work therefore provides physical insight in the coupled dynamics of a compliant ink channel with an entrained microbubble. 

\end{abstract}


\maketitle

\section{\label{sec:1} Introduction}
Inkjet printing is a successful industrial application of microfluidics that allows highly controlled contactless deposition of droplets at picoliter volumes.~\cite{Derby2010, Wijshoff2010,daly2015pharmaceutics,Lohse2021} Droplet formation is driven either by a heat-induced vapor bubble in thermal inkjet printing or by acoustic pressure waves in piezoacoustic inkjet printing.~\cite{Hoath2015,Dijksman2019} The higher versatility of piezoacoustic inkjet printing results from the wide range of inks that can be jetted, i.e., it is not limited to aqueous inks. As a result, piezoacoustic inkjet printing is widely used in an industrial setting for applications in document printing, graphic arts, and package printing.~\cite{Wijshoff2010} Furthermore, piezo driven inkjet technology has  been successfully employed for emerging applications in electronic device fabrication,~\cite{majee2016graphene, majee2017graphene, eshkalak2017cnt, vilardell2013ceramic, moya2017electrochemical, eggenhuisen2015solarcells, hashmi2015solarcells, shimoda2003displays, jiang2017displays} ceramics,~\cite{derby2015ceramic} pharmaceuticals,~\cite{daly2015pharmaceutics} and biomaterials.~\cite{simaite2016muscles, hewes2017microvessels, nakamura2005tissue, villar2013tissue}

Inkjet printing is a highly stable process through which billions of droplets, identical in volume and velocity, can be reproducibly jetted.~\cite{Fraters2019a,Derby2010, Wijshoff2010,daly2015pharmaceutics,Lohse2021} However, the jet stability can be compromised by the entrainment of one or more microbubbles into the ink channel.~\cite{deJong2006b, dejong2006entrapped, jeurissen2008effect, jeurissen2009acoustic, lee2009dynamics, kim2009effects, jeurissen2011regimes, vanderbos2011infrared, Fraters2019b} Bubbles were shown to nucleate on inkophobic dirt particles suspended in the ink.~\cite{Fraters2019a} Other potential entrainment mechanisms include cavitation inception in the rarefaction phase of the pressure wave~\cite{brennen1995cavitation} and bubble pinch-off from an inward gas jet formed at the oscillating air-ink meniscus driven by flow focusing~\cite{Fraters2021} or arising from instabilities associated with higher order meniscus modes.~\cite{vanderMeulen2020}  After entering the ink channel, the bubble is either jetted out within a few actuation cycles or forced  toward the channels walls by acoustic radiation forces,~\cite{Crum1975} where it will be trapped.~\cite{Fraters2019a} The trapped bubble has been shown to grow over hundreds of actuation cycles, due to rectified diffusion,~\cite{crum1984rectified, leighton1994acoustic, brennen1995cavitation, brenner2002single-bubble} before it reaches its maximum volume typically of the order of tens of picoliters (diameter of $\sim$25~$\mu$m). The compressibility of the bubble results in strong volumetric bubble oscillations in response to the driving acoustics in the ink channel. Consequently, a grown bubble suppresses the pressure buildup at the nozzle required for the ejection of a droplet, thus leading to nozzle failure.~\cite{Fraters2019a,vanderbos2011infrared} For this reason, bubble entrainment, growth, and translation in an operating piezo-driven ink channel have been extensively studied in order to develop mitigation strategies for nozzle failure.~\cite{Hine1991,Brock1984,DeJong2006,deJong2006b,vanderbos2011infrared,Fraters2019a,Lohse2021}

The dynamics of an acoustically driven microbubble is well described by the celebrated Rayleigh-Plesset equation.~\cite{Lauterborn1976} Linearization of the Rayleigh-Plesset equation shows that, in the limit of small amplitudes of oscillation, the acoustically driven bubble can be modeled as a simple mass-spring system where the inertia of the surrounding liquid (mass) oscillates against the compressible gas core (spring).~\cite{Leighton1994, Versluis2020} The corresponding bubble eigenfrequency is inversely proportional to its size, as was shown nearly a century ago by Minnaert.~\cite{Minnaert1933} 
 
Bubbles present in the ink channel can be detected by using the piezo not only as an actuator but also as a pressure sensor.~\cite{dejong2006entrapped,Kwon2007}  Once the piezo driving pulse has been applied, the ring-down characteristics of the resonating ink channel cavity can be measured by switching to a piezo-readout circuit. Thus, this pulse-echo technique can be applied to characterize changes in ink channel acoustics upon the entrainment of a bubble.~\cite{Kwon2009,dejong2006entrapped,vanderbos2011infrared,Fraters2019b}  The acoustics of today's ink channels---typically fabricated in silicon using MEMS technology---is well understood.~\cite{kim2014mems,Wijshoff2010} The ink channel length in these printheads is much smaller than the acoustic wavelength and therefore, the acoustics can be approximated by that of a Helmholtz oscillator.~\cite{dijksman1998hydro-acoustics,Wijshoff2010,Dijksman2019}  In essence, a Helmholtz oscillator is a mass-spring system where the mass is that of the ink in the nozzle and the restrictor and the spring constant is the sum of the compliance of the actuator and that of the ink volume in the chamber.

Recently Fraters~\emph{et al.}~\cite{Fraters2019b} showed that the resonance frequency of an ink channel increases above its Helmholtz resonance frequency when a bubble is entrained. Furthermore, they reported the exact same ink-channel resonance frequency for two very different bubble volumes, indicating that the acoustic signal measured by the piezo is not simply that of the bubble alone but, in fact---as we will show here---that of a coupled mass-spring system of the Helmholtz oscillator and the bubble.~\cite{Fraters2019b} These intriguing results and a lacking physical interpretation motivated the present work. 

Here, a simple lumped--element model is developed (Section~\ref{sec:3}) to describe the dynamics of the coupled mass-spring systems of the bubble and the ink channel, with the aim to gain physical insight into the resonance behavior of the coupled system.  To validate the model, we describe in Section~\ref{sec:2} how we measured the acoustic ring-down signal of MEMS-based ink channels as a function of the bubble volume and the number of bubbles. One of the printheads comprised a glass-nozzle plate, thus allowing ultra-high-speed optical imaging at a frame rate of 1~million frames/s of the radial bubble oscillations to obtain the microbubble resonance curve. The comparison between the model and experiments is presented and discussed in Section~\ref{S:4}. We end the paper with conclusions (Secion~\ref{S:conl}). 

\section{Modeling}\label{sec:3} 
We start this section by describing our lumped element model with coupled inertances and compliances (masses and springs, respectively) and we derive its eigenfrequencies (Section~\ref{S:2a}). Next, the compliance of a bubble (Section~\ref{S:Cb}) and the inertances in ink channel--bubble system are derived or modeled (Section~\ref{S:Inerts}) such that the eigenfrequencies can be calculated. Finally, the resonance curve of the entrained bubble is derived in Section~\ref{S:Bubble}.

\subsection{Eigenfrequencies of the lumped element model} \label{S:2a}

\begin{figure}[t]
\begin{center}
\includegraphics[width=.95\columnwidth]{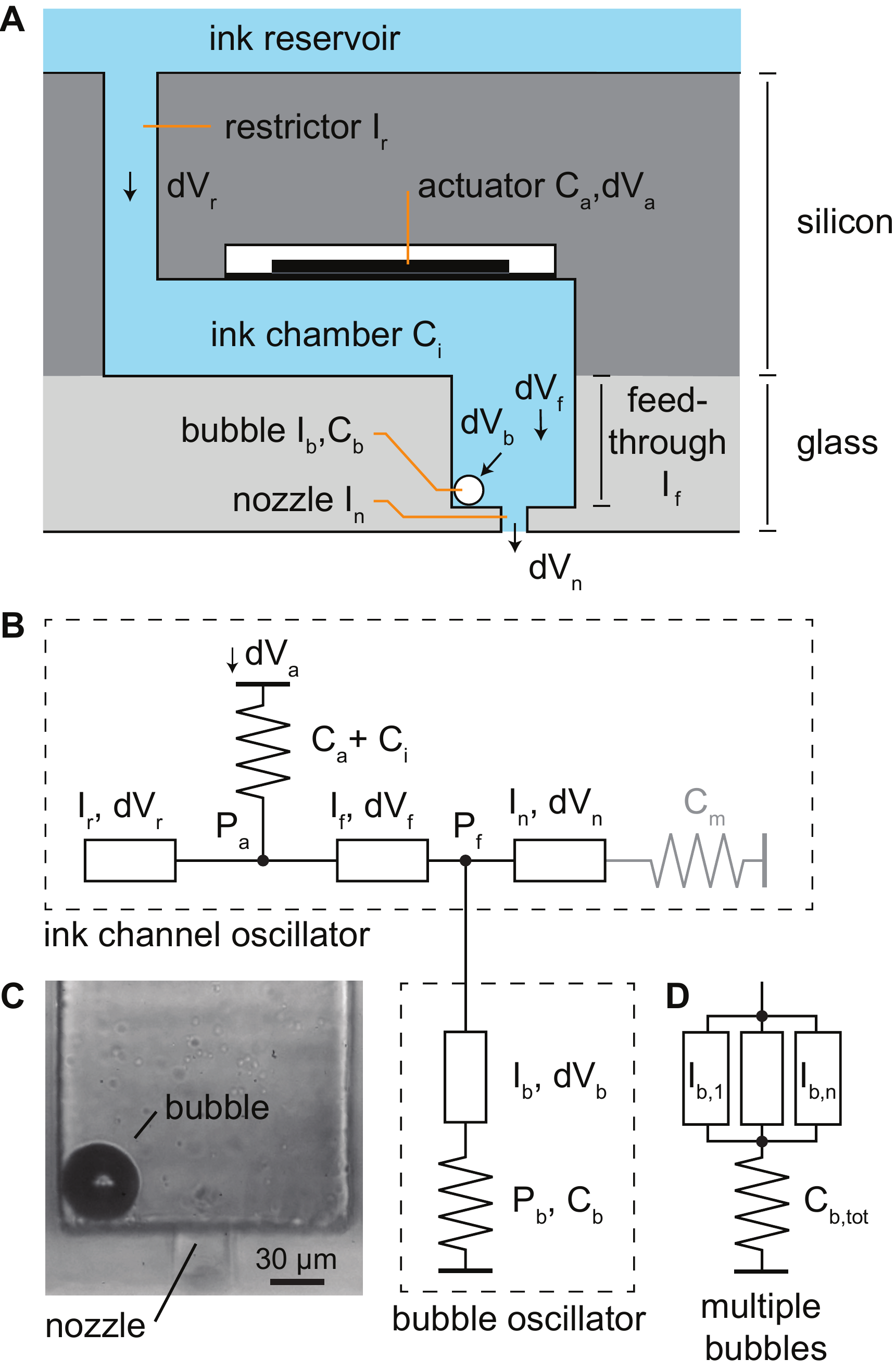}
\caption{(A) Schematic of the employed ink channel with the ink chamber, restrictor, feedthrough, nozzle, and entrained bubble. The ink channel is driven by a piezoelectric actuator deposited on a flexible membrane. (B) Lumped-element model (LEM) of the ink channel shown in (A) indicating the inertances of the restrictor $I_r$, the feedthrough $I_f$, the nozzle $I_n$, and the bubble $I_b$. The compliances of the actuator, the ink, the bubble, and the meniscus are indicated by $C_a$, $C_i$, $C_b$, and $C_m$, respectively. The pressure driving the system resulting from a volume change of the actuator $dV_a$ is given by $P_a$. This driving pressure is decreased to $P_f$ by accelerating the inertance of the feedthrough $I_f$. The pressure inside the gas core of the bubble is given by $P_b$. (C) Microscopy image showing the cylindrical feedthrough with an entrained bubble in the corner at its base. (D) Equivalent bubble-oscillator model circuit for the case of multiple entrained bubbles.}
\label{F:1}
\end{center}
\end{figure}

A schematic drawing of the ink channel employed in this work is shown in Fig.~\ref{F:1}A. The acoustic pressure can be assumed uniform within the ink chamber since its dimensions are much smaller than the acoustic wavelength $\lambda$. Due to the compliant actuator that constitutes the top wall of the ink chamber, the ink chamber changes volume in response to pressure changes. The volume of ink in the chamber adds another (minor) contribution to the total compliance of the system $C_t$ due to its finite speed of sound. Since the feedthrough is much smaller in volume than the ink chamber and since it has rigid walls, it has negligible compliance and is therefore accounted for in the present analysis as an incompressible mass (Fig.~\ref{F:1}B). The ink channel without a bubble is similar to a classical Helmholtz resonator, or mass-spring system, where the mass of the ink inside the restrictor oscillates out of phase with that in the feedthrough and the nozzle. When a bubble is present in the ink channel it adds another mass-spring system to the ink channel oscillator. In our model, the bubble oscillator is connected between the feedthrough and the nozzle, as shown in Fig.~\ref{F:1}B. As we will show later, the system is only lightly damped (Fig.~\ref{F:4}), i.e., the measured resonance frequencies are approximately equal to the eigenfrequencies and therefore, we neglect damping in the system. Furthermore, in our analysis, we neglect the compliance of the meniscus ($C_m$ in Fig.~\ref{F:1}B) that results from its surface tension since $C_m \gg C_t$. As such, we do not model the low frequency slosh-mode in which the inertances of the restrictor $I_r$, feedthrough $I_f$, and nozzle $I_n$ oscillate in phase.~\cite{Wijshoff2010} Indeed, the frequency of the slosh mode ($\sim$15~kHz) is typically one order of magnitude lower than the resonance frequencies of interest. In what follows, we derive the eigenfrequencies of the model system of Fig.~\ref{F:1}B. 

We start with the Newton-Laplace equation:
\begin{equation}\label{Eq:1}
c^2 = \frac{K}{\rho},
\end{equation}
which relates the speed of sound $c$ to the liquid density $\rho$ and the isentropic bulk modulus $K$~\cite{Blackstock2001a}:
\begin{equation}\label{Eq:2}
K = - V \frac{dP}{dV},
\end{equation}
with $dP$ the infinitesimal pressure change upon an infinitesimal volume change $dV$ of a cavity with volume $V$. From Eqs.~\ref{Eq:1} and~\ref{Eq:2} it follows that:
\begin{equation} \label{Eq:dP}
dP = - \frac{1}{C} dV,
\end{equation}
showing that the pressure rise in the ink chamber with total volume $V$ is proportional to its infinitesimal decrease in volume. In the proportionality constant $1/C$, $C$ is defined as the acoustic compliance $ C = - dV/dP = V(\rho c^2)^{-1}$ (volume change per unit pressure). For the ink chamber shown in Fig.~\ref{F:1}A the total acoustic compliance $C_{t}$ is the sum of the compliance of the volume of ink $V$ in the chamber and that of the flexible actuator. The total compliance of the ink channel employed in this work is $C_{t}$~= 5$\times$10$^{-20}$~m$^3$/Pa (or 5~pL/bar) as was obtained from investigations on the resonance behavior of experimental printheads (without bubbles) in comparison with full finite element simulations in Ansys of the same printheads~\cite{Wijshoff2010}.

Using Eq.~\ref{Eq:dP}, the pressure rise in the ink chamber $dP_a$ is given by:
\begin{equation} \label{Eq:dP2}
dP_a =  \frac{1}{C_{t}} (dV_r-dV_f),
\end{equation}
where $dV_r$ indicates the volume influx into the ink chamber from the restrictor and -$dV_f$ that from the feedthrough. Positive volume fluxes are indicated by the direction of the arrows in Fig.~\ref{F:1}A. 

A pressure rise in the ink channel accelerates the mass of ink in the restrictor in negative direction and that in the feedthrough in positive direction. For the mass of ink in the restrictor, Newton's second law reads: $dP_a A_r=-\rho L_r A_r \ddot x$, with $L_r$ the length of the restrictor, $A_r$ its cross-sectional area, and $\ddot x$ the acceleration of its mass ($\rho L_r A_r$). Note that $A_r \ddot x$ gives the acceleration of the displaced volume: $d \ddot V_r$, such that Newton's second law can be written in terms of a volume flow rate, as follows:
\begin{equation} \label{Eq:N2L}
dP_a = -  I_r d \ddot V_r,
\end{equation}
where the proportionality constant:
\begin{equation} \label{inertance}
I_r=\frac{\rho L_r}{A_r},
\end{equation} 
is the acoustic inertance of the restrictor. Equations~\ref{Eq:dP} and~\ref{Eq:N2L} are equated to obtain a second-order differential equation that describes the forced motion of the inertance of the restrictor, as follows:
\begin{equation} \label{Eq:VR}
 \frac{1}{C_t} (V_r-V_f) = - I_r \ddot V_r.
\end{equation}
Note that we have now omitted the use of the differential sign $d$ for ease of readability. 

Similarly, we obtain a differential equation for the motion of the inertance of the nozzle that is driven by the pressure in the ink chamber, $P_a$ (Eq.~\ref{Eq:dP}), minus the pressure lost over the inertance of the feedthrough, as follows: $P_f = P_a - I_f \ddot V_f$. Thus, the equation of motion for the nozzle reads:
\begin{equation} \label{Eq:VN}
 \frac{1}{C_{t}} (V_r-V_f) - I_f \ddot V_f = I_n \ddot V_n .
\end{equation}

The full non-linear dynamics of a bubble with volume $V_b$ in response to a time-varying acoustic pressure is well described by the Rayleigh-Plesset equation.~\cite{Leighton1994} Here, we consider linear bubble oscillations (verified a posteriori using high-speed imaging). Linear bubble oscillations are well described by a harmonic oscillator model where the spring constant is given by the compliance of the compressible bubble $C_b$ and the inertance $I_b$ by the mass of the fluid surrounding the bubble that is involved in the bubble oscillations. In Sections~\ref{S:Cb} and~\ref{S:numerics} we will further elaborate on $C_b$ and $I_b$. Here, we first focus on the equation of motion of the bubble: The inertance of the bubble is driven into motion by the pressure difference between the outside ($P_f$) and the inside of the bubble ($C_b^{-1} dV_b$), resulting in a third differential equation: 
\begin{equation} \label{Eq:VB}
 \frac{1}{C_{t}} (V_r-V_f) - I_f \ddot V_f - \frac{V_b}{C_b} = I_b \ddot V_b.
\end{equation}

The final equation that closes the problem follows from continuity:
\begin{equation} \label{Eq:D4}
\ddot V_f = \ddot V_n + \ddot V_b.
\end{equation}
Thus, in our model the inertances of the feedthrough, the nozzle, and the bubble are directly connected via the incompressible ink. 

The eigenfrequencies of the set of four coupled second-order differential equations (Eqs.~\ref{Eq:VR} -\ref{Eq:D4}) are found by assuming simple harmonic motion, $V(t)~=~V_A \sin(\omega t)$, with $V_A$ a constant amplitude (unimportant in the present derivation since we are interested in the eigenfrequencies) such that the system can be written as a set of coupled linear equations, as follows:
{\small 
\begin{equation} \label{Eq:M}
\begin{split}
\begin{bmatrix}
\omega^2 I_r - C_t^{-1} & C_t^{-1} & 0 &  0 \\
-C_t^{-1} & -\omega^2 I_f + C_t^{-1} & -\omega^2 I_n & 0 \\
-C_t^{-1} & -\omega^2 I_f + C_t^{-1} & 0& -\omega^2 I_b+ C_b^{-1} \\
0 & -\omega^2 & \omega^2 &\omega^2 
\end{bmatrix}
\times \\ 
\begin{bmatrix}
V_r \\
V_f \\
V_n   \\
V_b
\end{bmatrix} 
 = 0
\end{split}
\end{equation}}

\noindent The eigenfrequencies are then found by dividing the determinant of above 4$\times$4~matrix by $\omega^4$ (i.e. first four eigenfrequencies are zero) and by equating it to zero, as follows: 
\begin{equation} \label{Eq:Det}
\begin{split}
& - \omega^4 ( I_r I_f I_n  +  I_r I_f I_b  +  I_r I_n I_b ) \\
& + \omega^2 \left( \frac{I_f I_n}{C_t} + \frac{I_n I_r}{C_t} + \frac{I_f I_ r}{C_b} + \frac{I_n I_r}{C_b} + \frac{I_b (I_f+I_n+I_r)}{C_t} \right) \\
&  - \frac{I_r+I_f+I_n}{C_t C_b} = 0.
\end{split}
\end{equation}
This equation is quadratic in terms of $x=\omega^2$, as follows: $ax^2+bx+c=0$, with angular eigenfrequencies:
\begin{equation} \label{Eq:omega}
\omega_0 = \sqrt{ \left(-b\pm \sqrt{b^2-4ac} \right)/2a}.
\end{equation}
Thus, the coupled ink channel--bubble system has two eigenfrequencies $f_1$ and $f_2$ (if $b^2>2ac$, and $f=\omega/2\pi$). Note that $a$, $b$, and $c$ are already given in Eq.~\ref{Eq:Det}. However, for completeness they are provided here as well: $a = -\left( I_r I_f I_n  +  I_r I_f I_b  +  I_r I_n I_b\right)$, and $b= I_f I_n/C_t + I_n I_r/C_t + I_f I_ r /C_b + I_n I_r/C_b + I_b (I_f+I_n+I_r)/C_t $, and $c= - (I_r+I_f+I_n)/(C_t C_b)$. 

As is shown later, an entrained bubble can grow due to rectified diffusion into a large bubble with a much larger compliance than that of the actuator. Therefore, it is of interest to discuss the limiting case with infinite bubble compliance ($C_b\gg C_t$), by neglecting the term $-V_b/C_b$ in Eq.~\ref{Eq:VB}. By doing so, the eigenfrequency of the ink channel is simply that of a classical Helmholtz resonator:
\begin{equation} \label{Eq:H}
\omega_L  = \sqrt{\frac{1}{C_{t}} \left( \frac{1}{I_r}+\frac{1}{I_f + \left( I_n^{-1}+ I_b^{-1} \right)^{-1} }\right)},
\end{equation}
with two inertances (masses), i.e.,  the inertance of the restrictor $I_r$, and an effective inertance: the sum of the inertance of the feedthrough $I_f$ and that of the nozzle $I_n$ and bubble $I_b$ in parallel. Note that this can be directly inferred from Fig.~\ref{F:1}B, without writing out the equations.

Equation~\ref{Eq:H} also shows that the eigenfrequency of the ink channel without a bubble is given by:
\begin{equation} \label{Eq:HP}
\omega_H  = \sqrt{\frac{1}{C_t} \left( \frac{1}{I_r}+\frac{1}{I_f + I_n }\right)}.
\end{equation}
Above equation is the eigenfrequency of the classic Helmholtz oscillator ($\omega^2 = (C_t I)^{-1} $) with effective inertance $I$~=~${1/I_r}+1/(I_f + I_n) $.

\subsection{Compliance of a bubble} \label{S:Cb}
In order to calculate the eigenfrequencies of the coupled system (Eq.~\ref{Eq:omega}), an expression for the compliance of a bubble as a function of its volume $V_B=\frac{4}{3}\pi R_0^3$, is required. Using the thermodynamic relation $PV^\gamma =$~const., for a bubble of radius $R_0$ filled with an ideal gas it holds that:~\cite{Minnaert1933,Leighton1994}
\begin{equation} \label{Eq:CB1}
\begin{split}
\left(P_0+2\sigma/R_0\right) V_B^\gamma & = \\
 \left(P_0+2\sigma(\frac{4\pi}{3V_b})^{1/3} \right) V_b^\gamma & =  \text{constant},
\end{split}
\end{equation}
with $P_0$ the pressure outside the bubble, $\sigma$ the interfacial tension, and $\gamma$ the polytropic exponent. By taking the derivative of Eq.~\ref{Eq:CB1} with respect to $P$ and $V$ the compliance of a bubble is obtained, as follows:
\begin{equation} \label{Eq:CB2}
\begin{split}
C_b^{-1} = - \frac{dP}{dV} = \frac{\gamma P_0}{V_b}+ 2\sigma \left(\frac{4\pi}{3V_b^4}\right)^{1/3} \left(\gamma-\frac{1}{3}\right)
\end{split}
\end{equation}
Note that the compliance of a bubble without surface tension equals: 
\begin{equation} \label{Eq:CBfree}
C_{b,free}=\frac{V_b}{\gamma P_0}.
\end{equation} 
The latter equation holds for any arbitrarily shaped air-filled cavity with volume $V_b$, or for multiple air-filled cavities with a total volume $V_b$, and it is not affected by confinement nor the number of bubbles.

\subsection{Inertances of the system} \label{S:Inerts}
\subsubsection{\small {Nozzle, restrictor, and feedthrough}}
The inertances of the nozzle, restrictor, and the feedthrough can be calculated using Eq.~\ref{inertance}. However, Eq.~\ref{inertance} underpredicts the inertance due to the fact that in a tube with oscillatory flow, the liquid just outside the tube is non stationary and thereby takes part in the oscillation. A well known end correction for the inertance of a tube is known in literature as the piston approximation.~\cite{Ingard1953,Blackstock2001a} It makes use of the assumption that the velocity at the end of an open tube is constant over the cross section of the tube. With this approximation, the inertance of a tube with radius $R_t$ is increased from that in Eq.~\ref{inertance} by increasing the effective tube length by $\Delta L = 8R_t(3\pi)^{-1} \approx 0.85 R_t$. The exact problem of the inertance of a circular hole in a thin plate is treated by Landau and Lifshitz~\cite{Landau2013} (page 25, problem 1):
\begin{equation} \label{Eq:EndCor}
\Delta L =  \frac{1}{4} \pi R_t \approx 0.79 R_t,
\end{equation}
which is close to the result from the piston approximation.

\subsubsection{\small{Inertance of a confined bubble}} \label{S:numerics}
\begin{figure*}[t]
\begin{center}
\includegraphics[width=1\textwidth]{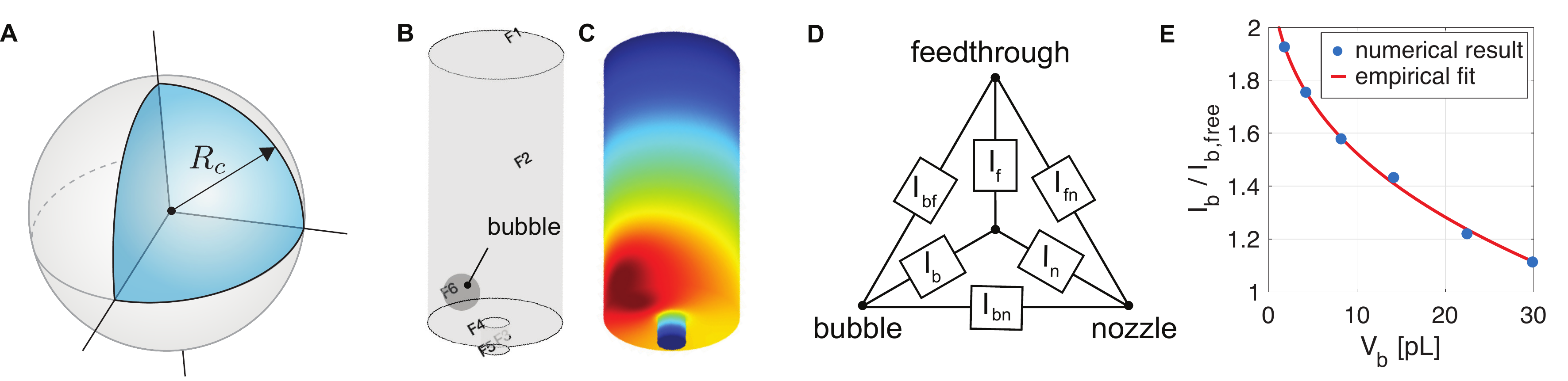}
\caption{(A) Schematic of a bubble in the corner of a channel with $1/8$ of the surface area and volume of a spherical bubble with radius $R_c$. In this configuration, the inertance increases by 4~times with respect to that of a bubble in the free field of the same volume. (B) Numerical simulation geometry of the feedthrough and nozzle including a bubble in the corner at the base of the feedthrough. (C) Typical example of a calculated velocity potential field for a potential difference $d \phi$ between the bubble surface (F6 in B) and the nozzle exit (F5 in B) and feedthrough entrance (F1 in B).  (D) Using the velocity potential, the flow rates can be obtained and converted to inertances through multiplication by $\rho d \phi$, which gives the inertances $I_{bf}$, $I_{fn}$, and $I_{bn}$. These can in turn be converted to the inertances of the bubble $I_b$, feedthrough $I_f$, and the nozzle $I_n$. (E) Numerically obtained inertance of a confined bubble $I_b$ in the vertex at the base of the feedthrough as a function of bubble volume $V_b$ and normalized to the inertance of a bubble in the free field $I_{b,free}$ (Eq.~\ref{Eq:Ibfree}).}
\label{F:2}
\end{center}
\end{figure*}

As we have shown, the compliance of a bubble is not affected by confinement (section~\ref{S:Cb}). However, as we will show now, its inertance is. We start by considering the inertance of an unconfined bubble. The most straightforward route to arrive at the inertance of a bubble in the free field is by starting at the inertance of a tube with arbitrary cross-sectional area $A$ of which the inertance is found by integrating over the tube length $L$, as follows: $I = \rho \int_0^L A(l)^{-1}dl$. Similarly, for a bubble in the free field the inertance $I_{b,free}$ is obtained by integrating the latter equation from $R_0$ to infinity, as follows:
\begin{equation} \label{Eq:Ibfree}
I_{b,free} =  \int_{R_0}^\infty \frac{\rho}{4\pi l^2} dl = \frac{\rho}{4\pi R_0}.
\end{equation}
Confinement of the bubble decreases the area $A(l)$ of the shells $A(l)dl$ through which ink flows to or away from the oscillating bubble and it thereby increases its inertance. The limiting case with maximum increased inertance with respect to a bubble in the free field is that of a bubble located in a 90$^\circ$ vertex of an infinitely large cube, de-wetting the channel walls with 90$^\circ$ contact angles between the gas and the walls, see Fig.~\ref{F:2}A. For this configuration, the radius of the confined bubble $R_c$ (Fig.~\ref{F:2}A) is twice that of a spherical bubble in the free field $R_f$ of the same volume. Using Eq.~\ref{Eq:Ibfree}, integrating from $2 R_f$ to infinity, and taking into account the 8~times reduced surface area of the vertex bubble with respect to that of a spherical bubble with radius $R_c$, it then follows that its inertance is 4~times larger than for a bubble of equal volume in the free field. 

Figure~\ref{F:1}C shows a sideview optical microscope image of an entrained bubble in the printhead employed in the present work. The image shows a nearly spherical bubble that does not de-wet the channel walls from which a inertance increase of less than 4~times is expected. However, the ink flow towards and away from the oscillating bubble in the ink channel is not only confined by the walls to which it adheres, but also by the opposing channel walls thereby increasing its inertance. On the other hand, the presence of the nozzle decreases the inertance of the bubble, as the ink can flow through the nearby nozzle, reducing the pressure at the bubble for a given flow rate. On top of that, an additional bubble size dependence due to the channel geometry is expected. All in all, the inertance of the confined bubble shown in Fig.~\ref{F:1}C cannot be simply calculated using geometrical arguments. 

In what follows, we numerically obtain an estimate of the inertance of a spherical bubble in the vertex of the circular feedthrough using potential flow theory, i.e., by assuming irrotational and incompressible flow. Under these assumptions, the velocity potential $\phi$ satisfies the continuity equation (Laplace's equation): $\nabla^2 \phi = 0$. Laplace's equation was solved in Matlab using the Partial Differential Equation Toolbox to obtain the velocity potential. The 3D geometry of the feedthrough and the nozzle including a bubble were constructed in Solidworks for bubble radii of 5.0, 7.5, 10.0, 12.5, 15.0, 17.5, and 20.0~$\mu$m. Figure~\ref{F:2}B shows the model geometry including a bubble with a radius of 20~$\mu$m located in the vertex at the base of the feedthrough. The diameter and length of the feedthrough were as in the experimental printhead, i.e. 140~$\mu$m and 470~$\mu$m, respectively. For each bubble size, the velocity potential was calculated for 2 cases with different Dirichlet boundary conditions. For case~1, the potential of the bubble surface (F6 in Fig~\ref{F:2}B) was set $d \phi$ (units of m$^2$/s) higher than that of the nozzle exit and the feedthrough entrance. For case~2, the potential of the feedthrough entrance (F1 in Fig~\ref{F:2}B) was set $d \phi$ higher than that of the bubble and the nozzle exit (F5 in Fig~\ref{F:2}B). A typical modeled velocity potential for the first case is shown in Fig.~\ref{F:2}C (color scale range equals $d \phi$). From case~1, the flow rate from the bubble to the nozzle exit $q_{bn}$ was obtained together with that from the bubble to the feedthrough entrance $q_{fb}$ by integrating the velocity ($\nabla \phi$) over the nozzle exit surface and over the cross section of the feedthrough inlet, respectively. Similarly, from case~2, the flow rate from the nozzle exit to the feedthrough $q_{fn}$ was obtained. The inverse of these flow rates multiplied by $\rho d \phi$ gives the effective inertances between the bubble and the feedthrough $I_{bf}$, the bubble and the nozzle $I_{bn}$, and the feedthrough and the nozzle $I_{fn}$, as indicated in the in the diagram shown in Fig.~\ref{F:2}D. The inertances of the feedthrough, bubble, and nozzle are also shown in Fig.~\ref{F:2}D. Note that indeed, the effective inertances obtained from the numerics can be related to the individual inertances of the feedthrough, nozzle, and bubble as defined in Fig.~\ref{F:1}. Therefore, note that:
\begin{equation}\label{Eq:P}
P = I_f + I_b = \left(\frac{1}{I_{bf}} + \frac{1}{I_{fn} +I_{bn}} \right)^{-1},
\end{equation}
meaning that the inertance $I_f + I_b$ equals $I_{bf}$ in parallel with $I_{fn}+I_{bn}$. Similarly, it holds that:
\begin{equation}\label{Eq:Q}
Q = I_f + I_n = \left(\frac{1}{I_{fn}} + \frac{1}{I_{bn} +I_{bf}} \right)^{-1},
\end{equation}
and
\begin{equation}\label{Eq:R}
R = I_b + I_n = \left(\frac{1}{I_{bn}} + \frac{1}{I_{bf} + I_{fn}} \right)^{-1}.
\end{equation}
Equations~\ref{Eq:P}, \ref{Eq:Q}, and \ref{Eq:R} contain the 3~unknown inertances $I_f$, $I_n$, and $I_b$, for which can be solved, as follows:
\begin{equation} \label{Eq:If_num}
I_f  = \frac{P-Q+R}{2},
\end{equation}
\begin{equation} \label{Eq:In_num}
I_b = \frac{P+Q-R}{2},
\end{equation}
\begin{equation} \label{Eq:Ib_num}
I_n  = \frac{-P+Q+R}{2}.
\end{equation}

To verify the numerical method, we obtained the inertance of the bubble in a vertex as in Fig.~\ref{F:2}A for a cube with dimensions much larger than $R_c$. Indeed, we recover the analytical result of a 4~times increased inertance of the vertex bubble with respect to a bubble with the same volume in the free field. 

To further validate the numerical approach, we compare the numerically obtained inertances of the feedthrough (Eq.~\ref{Eq:If_num}) and nozzle (Eq.~\ref{Eq:In_num}) to their analytical values obtained from Eq.~\ref{inertance} in which $L$ is increased by the end correction factor (Eq.~\ref{Eq:EndCor}). The numerically and analytically obtained inertances of the feedthrough are in excellent agreement, i.e., they are 2.45$\times$10$^7$~kg/m$^4$ and 2.49$\times$10$^7$~kg/m$^4$, respectively. The numerically and analytically obtained inertances of the nozzle are both 4.9$\times$10$^7$~kg/m$^4$. Thus, the simple numerical approach using potential flow is able to accurately predict the inertances of the nozzle and feedthrough. 

The numerically obtained inertance of the bubble located in the corner at the base of the feedthrough (Eq.~\ref{Eq:Ib_num}), normalized by that of a bubble of the same volume in the free field (Eq.~\ref{Eq:Ibfree}), is plotted in Fig.~\ref{F:2}E as a function of bubble volume. For small bubble volumes, the inertance of the confined bubble increases by as much as a factor of two with respect to that of a bubble in the free field. Notably, for increasing bubble volumes, the normalized inertance decreases. At the maximum bubble volume studied in this work (30~pL), the inertance of the confined bubble is only a factor of 1.1~higher than that of a bubble in the free field. This decrease results from the fact that for larger bubbles, the distance between the bubble interface and the nozzle decreases, thereby increasing the flow-rate towards the nozzle at a constant pressure difference. This in turn lowers the effective inertance. The solid red curve in Fig.~\ref{F:2}E shows the fit: $I_b/I_{b,free}= -6.93\times 10^4 R_b+2.45$, that we use in Section~\ref{R:1} to calculate the eigenfrequencies of the ink channel as a function of bubble volume. 

\subsubsection{\small Inertance of multiple bubbles}
Inspired by experiments (in Ref.~\cite{Fraters2019b} and in Section~\ref{S:multibub}), we discuss the change in eigenfrequency of the coupled system when multiple bubbles are present in the vertices at the base of the feedthrough with total gas volume $V_b$, as compared to that of the system with a single bubble of the same volume. The equivalent model diagram for this bubble oscillator that comprises multiple bubbles is shown in Fig.~\ref{F:1}D. As shown before, by neglecting surface tension, the compliance of a given gas volume $V_b$ is independent of the total number of bubbles that comprise this volume (Eq.~\ref{Eq:CBfree}). On the other hand, the effective total inertance of multiple bubbles with total gas volume $V_b$ decreases with respect to that of a single bubble with the same volume due to the increased surface area to volume ratio. Indeed, the effective total inertance of the parallel circuit (Fig.~\ref{F:1}D) of the inertances of $N$ bubbles is given by: 
\begin{equation} \label{Eq:IN}
I_{tot}=\left( \sum_{n=1}^N = 1/I_{b,n} \right)^{-1},
\end{equation} 
with $I_{b,n}$ the inertance of bubble $n$. In our experiments performed at constant driving conditions, the total gas volume $V_b$ was always the same, independent of the number of bubbles. Furthermore, when multiple bubbles were present, these always had the same size (Section~\ref{R:multiple}). At a total gas volume $V_b$, the equivalent radius of $N$ bubbles of the same size equals: $R_N = N^{-1/3}R_0$, with $R_0$ the radius of the spherical bubble with volume $V_b$. Using the latter expression for $R_N$ and Eqs.~\ref{Eq:Ibfree} and \ref{Eq:IN}, we find:
\begin{equation} \label{Eq:Icollective}
I_{tot} = \left( \sum_N  \frac{1}{N^{1/3} I_{R_0}} \right)^{-1} = \frac{1}{N^{2/3}} I_{R_0}.
\end{equation}
Thus, at a constant total gas volume, the effective inertance decreases with the number of bubbles that comprise this gas volume. Thus, at a constant gas volume, the eigenfrequency of the ink channel is expected to increase with the number of bubbles due to a decreased effective inertance. 

\subsection{Resonance curve of the confined bubble} \label{S:Bubble}
In this section, we obtain a differential equation for the volumetric bubble oscillation amplitude expressed solely in terms of $V_b$ and its derivatives with a driving pressure term expressed in terms of $P_a$ (Fig.~\ref{F:1}B). This is of interest because both variables can be measured experimentally, i.e., the oscillation amplitude can be measured through high-speed imaging and the relative acoustic driving pressure is given by the amplitude of the ring-down piezosignal. Therefore, Eqs.~\ref{Eq:dP2} and \ref{Eq:D4} are substituted in Eq.~\ref{Eq:VN} to obtain an expression for $\ddot V_n$, as follows: $\ddot V_n = (P_a -I_f \ddot V_b)/(I_n+I_f)$. The latter expression is substituted together with Eqs.~\ref{Eq:dP2} and~\ref{Eq:D4} in Eq.~\ref{Eq:VB}, to obtain: 
\begin{equation} \label{Eq:bub}
I_{b,eff} \ddot V_b + \delta_{tot}\omega_b \dot V_b +\frac{1}{C_b }  V_b =  P_A~ e^{i \omega t},
\end{equation}
Above equation is that of a damped harmonic oscillator driven at an acoustic pressure amplitude of: $P_A = P_a \left(1- \frac{I_f}{I_n+I_f}\right)$. The effective inertance $I_{b,eff}$ is given by: 
\begin{equation}\label{Eq:Ibeff}
I_{b,eff}=I_b  +  \frac{1}{I_f^{-1} + I_n^{-1}}.
\end{equation}
Note from Fig.~\ref{F:1} that indeed, the effective mass that oscillates with the bubble spring $C_b$ equals the inertance of the bubble $I_b$ added to the equivalent inertance of the feedthrough and nozzle in parallel ($(1/I_f + 1/I_n)^{-1}$, see Eq.~\ref{Eq:Ibeff}). An ad-hoc dimensionless damping term was added to Eq.~\ref{Eq:bub} to account for the total damping $\delta_{tot}$ of the system in order to ensure a finite oscillation amplitude at resonance. The eigenfrequency of the bubble harmonic oscillator is found by assuming simple harmonic motion: $V_b(t)~=~V_B e^{i \omega t}$, with $V_B$ a constant, resulting in a simple form of the eigenfrequency:
\begin{equation} \label{Eq:omegab}
\omega_{b} = \sqrt{ \frac{1}{C_b I_b} }.
\end{equation}
Note that an increased inertance due to e.g. bubble confinement decreases the eigenfrequency. The Minneart eigenfrequency of a bubble in the free field is obtained when Eqs.~\ref{Eq:CBfree} and \ref{Eq:Ibfree} are substituted in above equation:
\begin{equation}\label{Eq:Minneart}
\omega_{b,free} =  \frac{1}{R} \sqrt{ \frac{3\gamma P_0}{\rho} } ,
 \end{equation}
demonstrating the characteristic $1/R$ dependency. The resonance frequency $\omega_{res}$ is slightly lowered by the total damping, as follows: $\omega_{res} = \omega_b \sqrt{1-\delta_{tot}^2/2}$~\cite{Versluis2020}.

The complex volumetric oscillation amplitude $A$ for the steady-state response of the bubble harmonic oscillator is obtained by substituting the solution $V_b = A e^{i \omega t + \varphi}$ in Eq.~\ref{Eq:bub}, which can then be written in absolute form as follows:~\cite{Leighton1994} 
\begin{equation} \label{Eq:A}
|A| = \frac{P_A}{I_{b,eff}} \frac{1}{\sqrt{(\omega_b^2-\omega^2)^2+(\delta_{tot}\omega_b \omega)^2}}.
\end{equation}
For small amplitudes of oscillation, the radial oscillation amplitude is given by: $R_\epsilon \simeq |A| (4 \pi R_0^2)^{-1}$. Equation~\ref{Eq:A} plotted as a function of the driving frequency then gives the resonance curve of the bubble in the printhead. 

\section{Experimental methods}\label{sec:2} 

\subsection{Printhead and model ink}
To validate the model, experiments were performed on two different printheads, i.e. (i) an experimental printhead with an optically transparent nozzle plate (Fig.~\ref{F:1b}), and (ii) a printhead fully fabricated in silicon. We start by describing the experiments performed on the printhead with a glass nozzle plate of which the fabrication was described in detail by Fraters~\emph{et al.}~\cite{Fraters2019a} In short, the printhead had a silicon-based functional acoustic part to which an optically transparent 500~$\mu$m thick fused-silica nozzle-plate chip was bonded (see Fig.~\ref{F:1}A). In the fused-silica chip, the feedthrough and the nozzle were fabricated at a precision of 1~$\mu$m through femtosecond-laser-pulse printing and subsequent chemical etching (FemtoPrint, Muzzano, Switzerland). The nozzles were cylindrical, with a diameter and length of 30~$\mu$m. The feedthrough channels were cylindrical as well with a diameter of 140~$\mu$m and a length of 470~$\mu$m. 

The employed model ink was Decanol (Sigma Aldrich) which has a surface tension of 28~mN/m, a density of 830~kg/m$^3$, and a viscosity of 11~mPa$\cdot$s. The printhead was actuated by a trapezoidal driving pulse with a rise and fall time of 1~$\mu$s and a high time of 3.0~$\mu$s, optimized for the bubble-free ink-channel Helmholtz resonance frequency of 125 kHz (full width--half maximum of the pulse was half a period, see Fig.~\ref{F:1b}). The driving pulse was generated by an arbitrary waveform generator (Wavetek~195) and amplified to an amplitude of 25~V by an amplifier (Krohn-Hite 7602M). For the ink channel without an air bubble, at this driving voltage, the velocity of the jetted droplets was 6~m/s. 

\subsection{Resonance frequency of the ink channel as a function of bubble volume}
\begin{figure}[tb]
\begin{center}
\includegraphics[width=1\columnwidth]{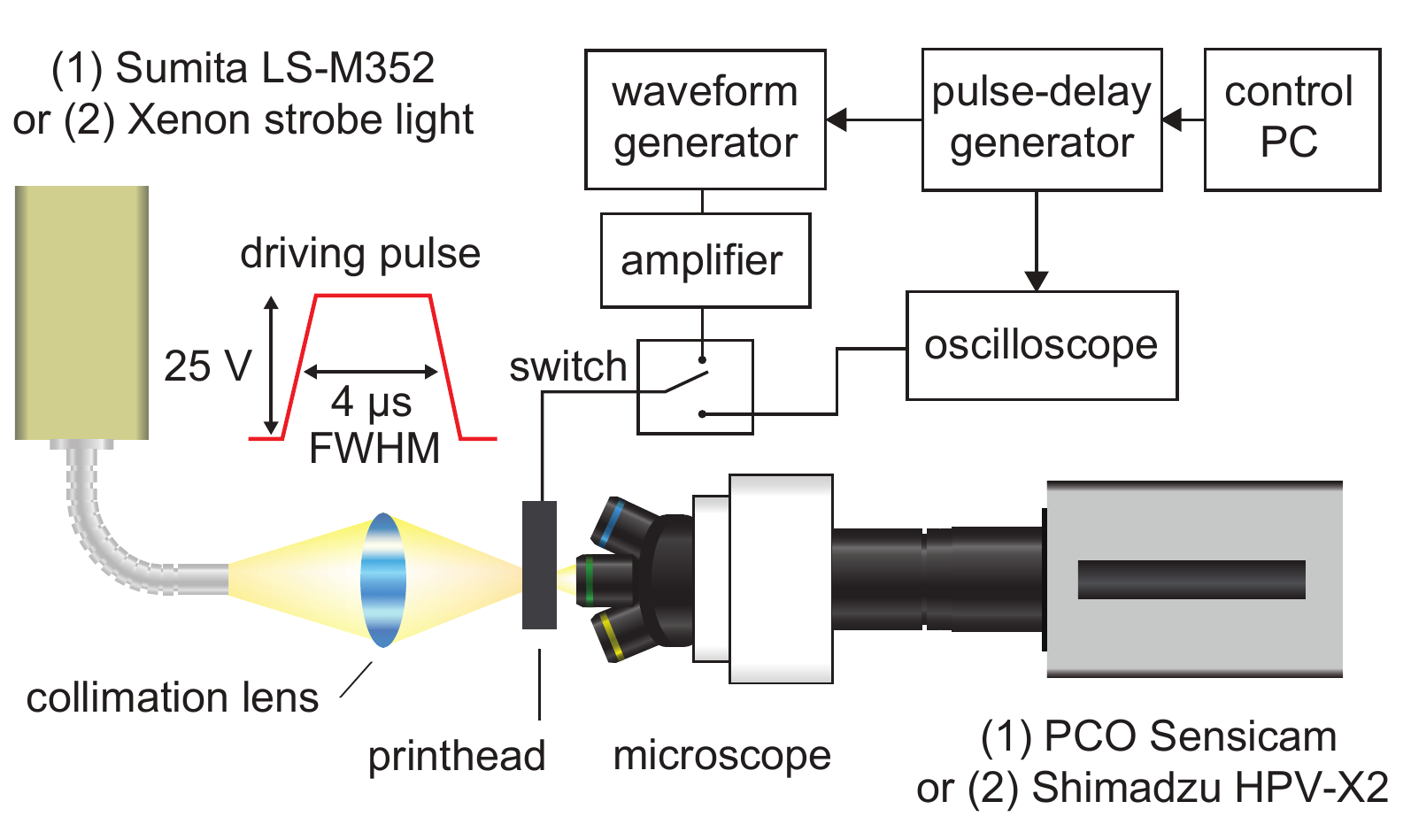}
\caption{Schematic representation of the experimental setup employed to study the resonance behavior of a inkjet channel with an entrained bubble.}
\label{F:1b}
\end{center}
\end{figure}

A bubble was   entrained stochastically into the ink channel by jetting droplets at a drop-on-demand (DOD) frequency of 20~kHz.~\cite{Fraters2019a} To this end, a pulse-delay generator (Berkeley Nucleonics Corp., BNC 575) was employed to trigger the arbitrary waveform generator (AWG) at 20~kHz. Once entrained, the bubble was allowed to reach its equilibrium size by rectified diffusion. Subsequently, the DOD frequency was decreased from 20~kHz to 2~Hz to allow the bubble to slowly dissolve from its equilibrium radius of typically 20~$\mu$m to its complete dissolution over the course of several minutes. During bubble dissolution, both the ring-down of the piezosignal and the bubble size were measured at the DOD frequency of 2~Hz. 10~$\mu$s before the start of the piezo actuation pulse, a camera (PCO Sensicam QE Dual Shutter) was triggered to capture an image of the bubble. A continuous light source (Sumita LS-M352) was used to back-illuminate the printhead, see Fig.~\ref{F:1b}. As the image was captured before the piezo was actuated, the bubble was at rest and motion blur-free images were captured at an exposure time of 10~$\mu$s. The camera was connected to a modular microscope (BXFM-F, BXFM-ILHS, Olympus) equipped with a 20$\times$ magnifying objective (SLMPLN20x, Olympus) resulting in an optical resolution of 365~nm/pixel. After image acquisition, the piezo was actuated to drive the ink channel acoustics. When the piezo voltage decreased below 1~V, a home made electronic circuit switched the electronic piezo connections automatically to an oscilloscope (Tektronix TDS5034b), via an in-house developed current amplifier. The oscilloscope was operated in sequence mode, i.e., at every received trigger from the pulse-delay generator, it stored a single piezo ring-down signal in its sequential memory. One measurement typically consisted of 500~bubble images and their corresponding ring-down piezosignals. 

The  ring-down piezosignals were stored on a personal computer and processed offline in Matlab to obtain their amplitude spectra. From the amplitude spectra, the frequency at maximum amplitude, or resonance frequency of the coupled bubble-ink channel oscillator, was measured and plotted against the corresponding bubble volume. The bubble volume was obtained by processing the captured images using a custom-made image processing script programmed in Matlab.~\cite{Segers2014} First, the center of the bubble was found by binarizing the images using an intensity threshold. The images were then converted to polar coordinates and the radial intensity lines starting from the bubble center were then averaged over all angles to obtain an average intensity profile. The bubble size was obtained from the inflection point of this intensity profile.~\cite{Segers2014}

\begin{figure*}[htb]
\begin{center}
\includegraphics[width=.95\textwidth]{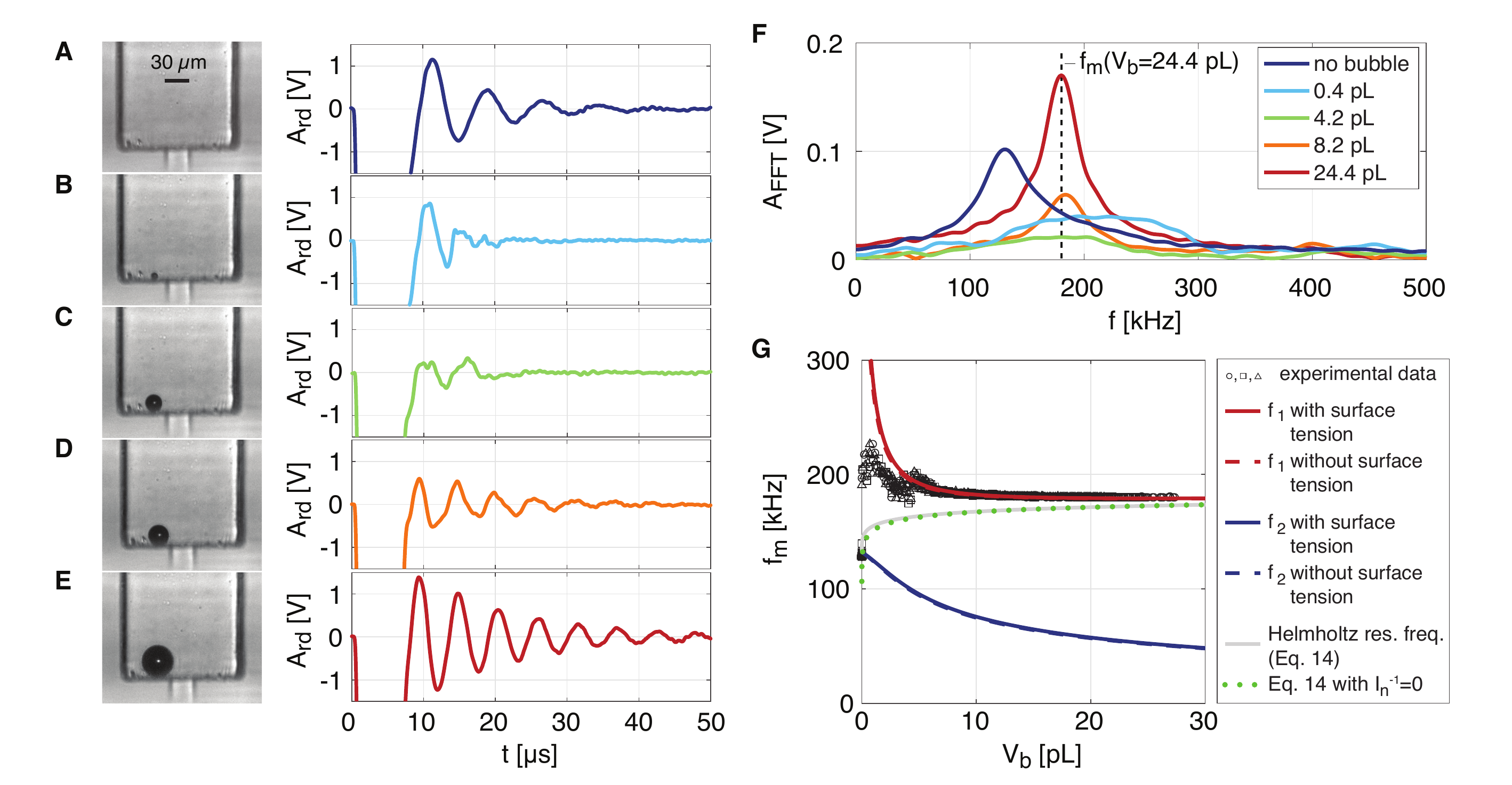}
\caption{(A) Microscope image of the feedthrough and nozzle without a bubble and the corresponding amplitude of the ring down piezosignal $A_{rd}$ as a function of time. (B)-(E)  Microscope images and corresponding ring down piezosignals for bubble volumes of 0.4, 4.2, 8.2, and 24.4~pL, respectively. (F) Amplitude spectrum of the ring down signals shown in (A)-(E). (G) Measured most dominant frequency $f_{m}$ of the coupled ink channel--bubble system as a function of bubble volume $V_b$ for three repeated experiments (open symbols). The two modeled eigenfrequencies ($f_{1}$ and $f_{2}$) of the coupled system are plotted as the red and blue solid curves. The dashed curves show the modeled eigenfrequencies for a bubble without surface tension. The solid gray line shows the eigenfrequency of the ink channel in the limit of an infinite bubble compliance (Eq.~\ref{Eq:H}). The green dotted curve shows the limit of infinite bubble compliance and infinite nozzle inertance ($I_n$ Eq.~\ref{Eq:H}) resembling a feedthrough exit open to the atmosphere. }
\label{F:4}
\end{center}
\end{figure*}

\subsection{Resonance frequency for multiple bubbles} \label{S:multibub}
In the second experiment, we measured the resonance frequency of an ink channel with multiple bubbles. These experiments were performed on the printhead with a feedthrough with a rectangular cross-section with rounded corners (Fig.~\ref{F:5}). In this experiment, the entrained bubbles were fissioned at a high driving voltage ($>$~35~V). The rectangular cross-section of the feedthrough resulted in the translation of the bubble fragments into the different corners of the feedthrough. Bubble fission and subsequent translation of the bubbles into different corners was a stochastic process which allowed for the study of different configurations with either 1, 2, 3, or 4 bubbles present, simply by repeating the experiment. After bubble fragmentation and translation, the driving conditions of the piezo were changed to a voltage amplitude of 25~V and a DOD frequency of 20~kHz to maintain a stable size for all bubbles (rectified diffusion and dissolution are in equilibrium).  This  printhead was opaque to visible light since it was fully fabricated in silicon. Therefore, imaging of the bubbles was performed using an infrared imaging setup, as described by Fraters~\emph{et al.}~\cite{Fraters2019b} To capture all bubbles in focus, imaging was performed in a plane parallel to the nozzle plate, i.e., from the bottom of the printhead. The ring-down piezosignals and bubble images were processed as before to extract the resonance frequency of the ink channel as a function of the number and volume of the bubbles.

\subsection{Resonance curve of a bubble}
The aim of the third and final experiment was to obtain a resonance curve of the bubble oscillations in the ink channel and to compare the experimental curve to the theoretical resonance curve obtained using Eq.~\ref{Eq:A}. This experiment therefore renders another validation of the model, and in particular, a validation and quantification of the numerically obtained inertance of the confined bubble (section~\ref{S:numerics}). 

The amplitudes of oscillation of a bubble in the optically transparent ink channel were measured as a function of the bubble size using a high-speed camera (Shimadzu HVP-X2) operated at 1~million frames/s. Illumination was provided by a Xenon flash-light with a pulse duration of 35~$\mu$s (Fig.~\ref{F:1b}). After stochastic bubble entrainment and growth by rectified diffusion at a DOD frequency of 20~kHz, the piezo driving was stopped. Subsequently, the flash light source, high-speed camera, oscilloscope, and waveform generator were triggered at the appropriate delays to record the bubble oscillations and the corresponding ring-down piezosignal. The experiment was repeated 41~times for bubble radii ranging from 6 to 20~$\mu$m. The bubble size was varied by allowing the bubble to dissolve for certain times before imaging was performed. Since the bubbles were oscillating non-spherically (Fig.~\ref{F:5}A), the equivalent time-dependent bubble radius was obtained from each frame of the high-speed recordings by measuring the area of the bubble from a binarized image after background subtraction. The amplitude spectrum of the radius versus time curves during the ring-down period was obtained through an FFT in Matlab while the equilibrium bubble size was first subtracted from the $R(t)$ curve. Finally, the amplitude of oscillation measured from the maximum amplitude in the amplitude spectrum was scaled by the amplitude of the ring-down piezosignal (at the same frequency) and plotted as a function of the bubble radius to obtain the resonance curve. Rescaling was done as the amplitude of the ring-down piezosignal is a relative measure of the acoustic driving signal in the ink channel ($P_A$ in Eq.~\ref{Eq:A}) and the piezo is linear at the frequencies of interest since its resonance frequency is $\sim$~2~MHz (as measured using a Laser Doppler vibrometer). 

\section{Experimental model validation and discussion} \label{S:4}

\subsection{Resonance frequency of the ink channel as a function of bubble volume} \label{R:1}
We now calculate the eigenfrequencies of the coupled ink channel--bubble system as a function of bubble volume by using the  numerically obtained inertances of the bubble, feedthrough, and nozzle and Eq.~\ref{Eq:omega}. Figure~\ref{F:4}A shows a microscope image of the optically transparent feedthrough without a bubble and the corresponding undisturbed ring-down piezosignal. Figures~\ref{F:4}B-E show images of a bubble in the feedthrough with a radius of  4.7, 10.0, 12.5, and 18.0~$\mu$m corresponding to volumes of 0.4, 4.2, 8.2, and 24.4~pL, respectively. The bubble is indeed located in the vertex at the base of the feedthrough exactly as in Fig.~\ref{F:1}C, however, it appears closer to the nozzle from the perspective of the observer. The time dependent amplitude $A_{rd}$ of the ring-down piezosignals corresponding to the different bubble sizes are shown in the column next to the microscope images. These ring-down signals show that once a bubble is entrained, the channel acoustics dramatically changes, i.e., the frequency and amplitude of the ring-down signals change. The amplitude spectrum $A_{FFT}(f)$ of the ring-down signals is shown in Fig.~\ref{F:4}F. Note that even though a bubble has a resonance frequency that is inversely proportional to its size, the frequency of the ring-down signals corresponding to the bubble volumes of 8.2 and 24.4~pL are the same. Indeed, Fig.~\ref{F:4}G shows that the measured most dominant frequency $f_{m}$ of the ring-down signal is nearly constant above bubble volumes $V_b$ of 8~pL (open symbols). 

The explanation can be found if we plot the two eigenfrequencies predicted by the model (Eq.~\ref{Eq:omega}), see Fig.~\ref{F:4}G (solid red and blue curves). In the modeled eigenfrequencies, the inertance of the bubble was that of a free bubble (Eq.~\ref{Eq:Ibfree}) multiplied by the numerically obtained correction factor (Fig.~\ref{F:2}E): $I_b/I_{b,free}= -6.93\times 10^4 R_b+2.45$. The first modeled eigenfrequency (solid red curve) is in excellent agreement with the measured most dominant frequency of the ring-down piezosignal. The solid curves show the predicted eigenfrequencies for a bubble with surface tension (Eq.~\ref{Eq:CB2}) and the dashed lines show those for a bubble without surface tension (Eq.~\ref{Eq:CBfree}). From the collapse of both curves it can be concluded that the Laplace pressure has a negligible effect on the dynamics of the coupled ink channel--bubble system. The second eigenfrequency of the coupled ink channel--bubble system (blue curves)  matches the measured resonance frequency of the ink channel without a bubble. Note that three data sets are plotted in Fig.~\ref{F:4}G (open symbols). These three data sets are typical examples of many repeated experiments showing that the frequency response of the ink channel versus the bubble volume is highly reproducible and that no mode switching was observed. From the excellent agreement of the modeled eigenmode with the data it can be concluded that the simple numerical modeling approach presented in this work accurately predicts the inertance of a confined bubble. 

\begin{figure}[htb]
\begin{center}
\includegraphics[width=.9\columnwidth]{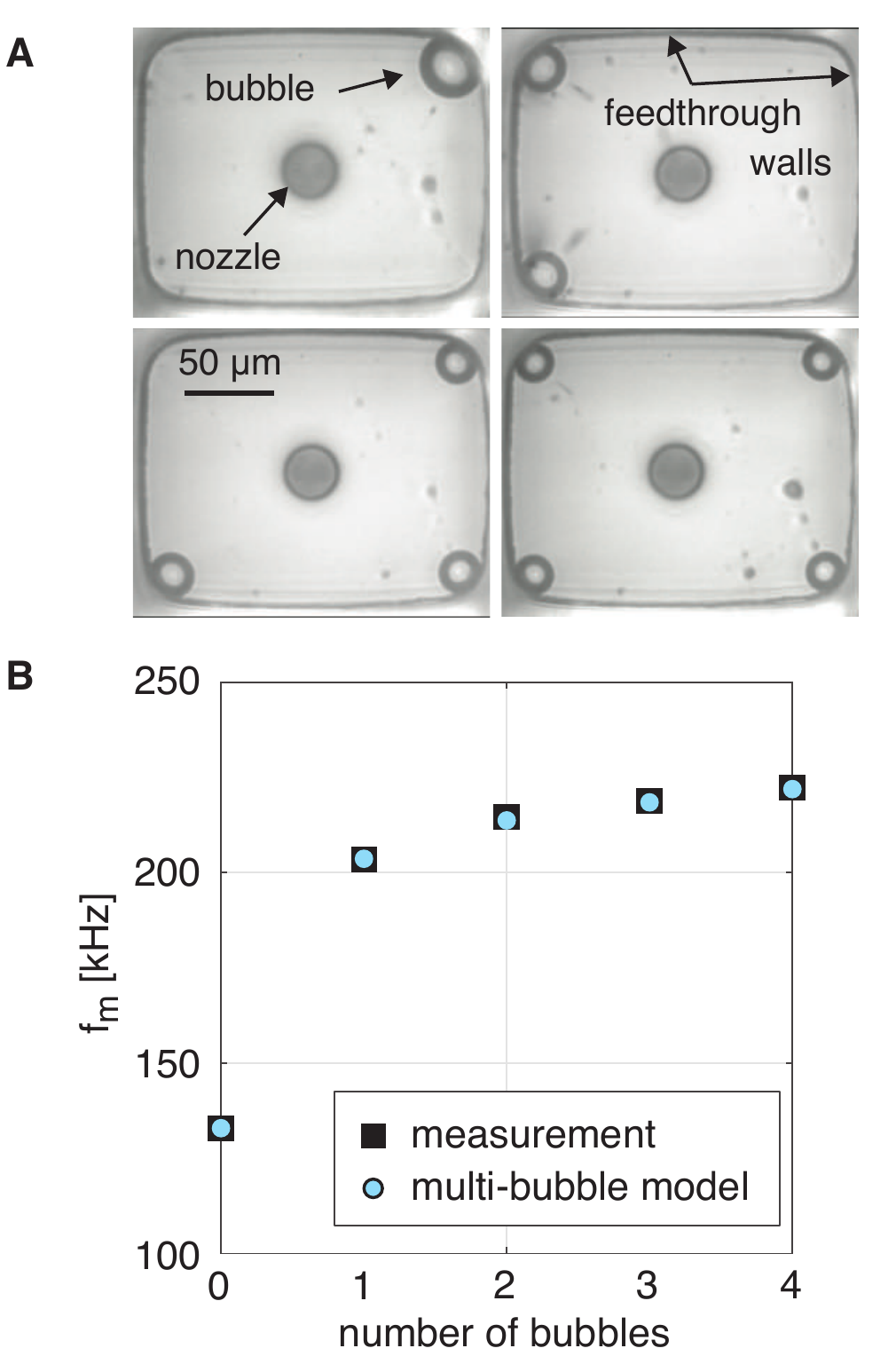}
\caption{Bottom view images of the silicon printhead obtained through infrared imaging. The dark circle in the center of the feedthrough is the nozzle. The resonance behavior of the ink channel either without a bubble, or with 1, 2, 3, or 4 bubbles was studied. (B) Measured most dominant frequency of the ring down piezosignal (resonance frequency) at a constant total gas volume of 20~pL and as a function of the number of bubbles (black squares). The modeled eigenfrequency of the ink channel is represented by the light-blue dots. The eigenfrequency was modeled using Eq.~\ref{Eq:omega} with an effective inertance that decreased with the number of bubbles according to Eq.~\ref{Eq:Icollective} thereby resulting in an increase in eigenfrequency with the number of bubbles.}
\label{F:6}
\end{center}
\end{figure}

The solid gray curve in Fig.~\ref{F:4}G shows the solution of Eq.~\ref{Eq:H}. It represents the modeled eigenfrequency of the coupled ink channel--bubble system in the limit of an infinite compliance of the bubble. The solid red curve and the gray curve approach the same asymptote showing that the eigenfrequency of the coupled ink channel--bubble system reaches a constant value for large bubble volumes due to the compliance of the bubble being much larger than that of the ink channel, and due to the inertance of the bubble being much lower than that of the nozzle. The latter statement is supported by the green dotted curve that shows the solution of Eq.~\ref{Eq:H} where the term $I_n^{-1}$ is set to zero. Thus, for large bubble volumes the base of the feedthrough effectively becomes an open end, i.e., the nozzle is decoupled from the system by the large compliance of the bubble which results in a nearly constant eigenfrequency of the coupled system that is independent of the bubble volume. 

The slight deviation between model and experiment as found for bubble volumes below 5 pL may result from nonlinear effects, both in terms of nozzle inertance and bubble dynamics. Nozzle inertance varies over the oscillation period due to the oscillatory changes in ink volume present in the nozzle. Once the nozzle starts to jet, these variations become highly nonlinear which thereby also changes the resonance frequency of the ink channel. Indeed, from the high-speed recordings an increased meniscus retraction was observed for the smaller bubble volumes suggesting that nonlinear nozzle inertance plays a role for small bubble volumes. Regarding bubble dynamics, at increasing acoustic driving pressure, the onset for nonlinear bubble dynamics is reached first when the bubble is driven near its resonance frequency. As we will show in Section IV-C, resonance occurs for bubbles with a radius of 8--10~$\mu$m, corresponding to bubble volumes of 2--4~pL. Figure~4G shows that for those bubble volumes $f_m$ deviates from the linear model, suggesting that bubble resonance plays a role. However, nonlinear phenomena were outside the scope of the present work that was aimed at developing a simple model for an ink channel with and without a bubble. Future work based on finite element modeling of the full nonlinear bubble-ink channel system may shed light on the role of nonlinearities on the resonance frequency.

\subsection{Resonance frequency change for multiple of bubbles} \label{R:multiple}

Figure~\ref{F:6}A shows bottom view images of the silicon printhead obtained through infrared imaging. Four different configurations were studied with either 1, 2, 3, or 4 bubbles. The bubbles were always located in a vertex at the base of the feedthrough as a result of the secondary Bjerknes force~\cite{Crum1975,Leighton1994} that is maximum in a corner with 3 pressure reflecting walls. Remarkably, at the employed driving condition, the total gas volume was the same for all four configurations, i.e., it was 20~pL $\pm$~3~pL, independent of the number of bubbles. The most dominant frequency of the ring down piezosignal $f_{m}$ was obtained as before and it is plotted in Fig.~\ref{F:6}B as a function of the number of bubbles (black squares). The modeled eigenfrequency for a total gas volume of 20~pL ($R_0$~=~17~$\mu$m) and a total inertance calculated using Eq.~\ref{Eq:Icollective}, with $I_{R_0}=1.3 I_{b,free}$ (Fig.~\ref{F:2}E), is also plotted in Fig.~\ref{F:6}B (red dots). The measured and modeled eigenfrequencies are in excellent agreement showing that indeed, the total inertance $I_{tot}$ of the system with multiple spherical bubbles decreases and that it is accurately represented by the equivalent circuit introduced in Fig.~\ref{F:1}D.

The good agreement of the lumped element model with experiment suggests that bubble-bubble interactions are negligible. The role of bubble-bubble interactions on $f_m$ can be studied by including a separate mass-spring system for each bubble and by subdividing the bubble inertance into two components: one for the flow toward the ink chamber and one for that toward the other bubbles(s). The model can be further extended toward nonlinear dynamics by performing a second-order linearization.~\cite{Prosperetti1974,Church1995,Sijl2011} 

\begin{figure}[htb]
\begin{center}
\includegraphics[width=1\columnwidth]{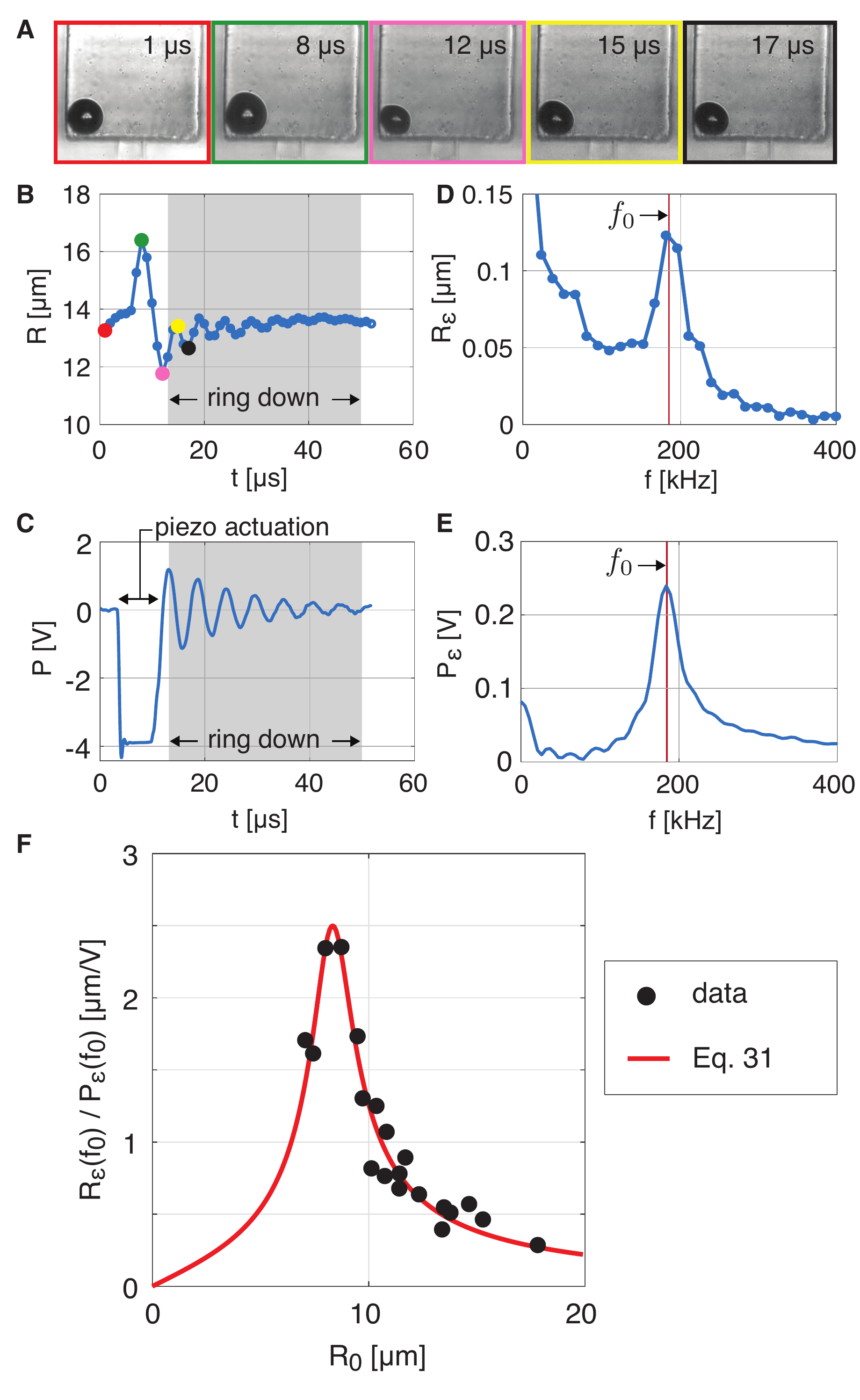}
\caption{(A) Selected frames of a high-speed recording of an oscillating bubble with an equilibrium radius of 13.6~$\mu$m. (B) Radius-time curve of the bubble in (A). (C) Ring down piezosignal measured simultaneously with the high-speed recording. (D) and (E) Amplitude spectra of the $R(t)$ curve and the ring down piezosignal over the ring down period (gray shaded area), respectively. The frequency of maximum response $f_0$ was the same for both the bubble dynamics and the ring down piezo signal. (F) Microbubble resonance curve, i.e., radial amplitude of oscillation at the frequency of maximum response $R_\epsilon(f_0)$ normalized by the amplitude of the ring down piezosignal also at the frequency of maximum response $P_\epsilon(f_0)$ as a function of the equilibrium bubble radius $R_0$.}
\label{F:5}
\end{center}
\end{figure}

\subsection{Microbubble resonance curve}
The resonance curve of a bubble in the ink channel was obtained by measuring the radius-time curves of 41~bubbles with equilibrium radii $R_0$ ranging from 6 to 20~$\mu$m. Selected frames from a typical ultra-high-speed recording of a bubble with an equilibrium radius of 13.6~$\mu$m are shown in Fig.~\ref{F:5}A.  The equivalent bubble radius over time $R(t)$ is plotted in Fig.~\ref{F:5}B. The color of the data points corresponds to time points indicated by the color of the frames around the images in Fig.~\ref{F:5}A. The ring-down piezosignal $P(t)$ measured simultaneously with  the bubble oscillations is plotted in Fig.~\ref{F:5}C. The piezosignal is clipped at $-4$~V for the duration of the piezo actuation pulse. The ring-down period is indicated by the gray shaded areas in Figs.~\ref{F:5}B and C. The amplitude spectrum of the radius-time curve during the ring-down period is shown in Fig.~\ref{F:5}D and that of the piezosignal is shown in Fig.~\ref{F:5}E. Note that the radial amplitude of oscillation $R_\epsilon$ over the ring-down period is at maximum at the same frequency as the dominant frequency in the ring-down piezosignal (indicated by $f_0$ in Figs.~\ref{F:5}D and E) confirming that the assumption of simple harmonic motion of the model is correct. $R_\epsilon$ at $f_0$ is normalized by the amplitude of the piezosignal $P_\epsilon(f_0)$ to obtain the resonance curve of bubble oscillations in the ink channel, see Fig.~\ref{F:5}F. Note the distinct resonance peak for a bubble with a radius of 8~$\mu$m.

The theoretical resonance curve was obtained by, first, calculating the eigenfrequency of the bubble using Eq.~\ref{Eq:omegab} in which the effective inertance given by Eq.~\ref{Eq:Ibeff} was used. In the calculation of the effective inertance, the numerically obtained inertances of the bubble (Fig.~\ref{F:2}E), nozzle, and feedthrough were used. Second, the radial amplitude of oscillation was calculated as a function of driving frequency using Eq.~\ref{Eq:A}. The resulting theoretical resonance curve is plotted in Fig.~\ref{F:5}F (solid red line) and it is in excellent agreement with the experimental data. To plot the theoretical curve, two parameters were tuned: the nondimensional damping coefficient $\delta$ was set to 0.25, and the amplitude was scaled due to the fact that we did not have acces to the absolute acoustic driving pressure ($P_A$ in Eq.~\ref{Eq:A}). However, most importantly, the position of the maximum was not fitted to the data and its excellent correspondence to the data again highlights the correctness of the numerically obtained inertance of the bubble. To substantiate this statement, note from Eqs.~\ref{Eq:Minneart} that the resonant bubble radius at the typical ultrasound frequency in the ink channel of 190~kHz is $\sim$~15~$\mu$m, whereas here, in the confinement of the ink channel it is $\sim$~8~$\mu$m. We foresee that this simple approach to obtain the inertance of a confined bubble is also relevant for the modeling of bubble dynamics outside the scope of inkjet printing, e.g., for the modeling of bubble dynamics in microfluidic devices and in the medical context of bubble oscillations confined by blood vessels.~\cite{Oguz1998,Segers2014, Versluis2020, Garbin2007}

The total nondimensional damping coefficient for an unconfined 8-$\mu$m radius free gas bubble in Decanol is approximately 0.7.~\cite{Devin1959, Versluis2020} The lower damping constant of 0.25 most likely results from the coupling of the bubble dynamics with the channel resonator. The relatively low damping suggests that minor damping contributions can be expected from the oscillatory flow in the feedthrough and nozzle and from the dissipation in the oscillating actuator membrane. Furthermore, the low damping constant justifies that the measured resonance frequencies (frequency of maximum response) can be directly compared to the modeled undamped eigenfrequencies owing to the small difference between the two (1.6\%). This observation also justifies the initial choice of neglecting damping in the present analysis.

\section{Conclusions} \label{S:conl}
The resonance behavior of a compliant piezo-driven inkjet channel with an entrained microbubble can be accurately described by solving the eigenfrequencies of a coupled system of two harmonic oscillators, i.e., one of the compliant ink channel and one of the bubble. The resonance frequency of the coupled ink channel--bubble system approaches a constant value when the compliance of the bubble dominates over that of the actuator. When the total gas volume is distributed over multiple entrained bubbles, the resonance frequency of the coupled system increases due to a decreased effective inertance of the bubbles, resulting from an increased bubble surface area to volume ratio. To describe the eigenfrequency of the coupled system, it is important to account for the relative change of inertance of the entrained microbubble with respect to that of a bubble in the free field due to its confinement by the ink channel. We demonstrate that the inertance of the confined bubble, and that of any other geometry such as a nozzle, can be accurately obtained using a simple potential flow approach. The present work therefore provides physical insight into confined bubble dynamics and provides a simple tool to model the inertance of a confined bubble, a nozzle, or any other complex geometry. The implications of the present work therefore reach beyond inkjet printing, and include research areas such as microfluidics, cavitation, and biomedical acoustics.

\section{Acknowledgements}
Financial support from an Industrial Partnership Programme of the Netherlands Organisation for Scientific Research (NWO), co-financed by Canon Production Printing, University of Twente, and Eindhoven University of Technology, and from the University of Twente. Furthermore, T.S. acknowledges financial support of the Max Planck Center Twente. We thank Uddalok Sen and Guillaume Lajoinie for stimulating discussions. 


%

\end{document}